\documentclass[a4paper,11pt]{article}
\usepackage{pos}
\usepackage{orcidlink}

\title{Evaluation of Feynman integrals via numerical integration of differential equations}

\author*{Pau~Petit~Ros\`as\,\orcidlink{0009-0009-8824-5208}\,}

\affiliation{Department of Mathematical Sciences, University of Liverpool,\\
Liverpool L69 3BX, 
U.K.}

\emailAdd{paupetit@liverpool.ac.uk}

\abstract{We revisit the idea of numerically integrating the differential form of Feynman integrals. With a novel approach for the treatment of branch cuts, we develop an integrator capable of evaluating a basis of master integrals in double and quadruple precision, with significantly smaller run times than other tools. This opens the door to evaluating higher complexity Feynman integrals on the fly in Monte Carlo generators, and enables a cheaper and easy to parallelise generation of grids for the topologies with prohibitive computational times. To show its performance, we test one- and two-loop integral families, achieving evaluation times in double precision of milliseconds and hundreds of milliseconds, respectively. We comment on the results and suggest room for improvement.}

\FullConference{17th International Symposium on Radiative Corrections: Applications of Quantum Field Theory to Phenomenology (RADCOR2025)\\
5-10 October 2025\\
Puri, India\\}

\begin{document}
\maketitle

\section{Introduction}
A dominant bottleneck in scattering-amplitude computations is the evaluation of multi-loop Feynman integrals, in particular for multi-scale processes with massive states~\cite{PetitRosas:2025xhm,Becchetti:2025qlu,Badger:2025ljy}. Among available methods, differential equations provide a systematic framework~\cite{Kotikov:1990kg,Gehrmann:1999as}. After integration-by-parts (IBPs) reduction~\cite{Chetyrkin:1981qh,Laporta:2000dsw}, a vector of master integrals (MI) obeys a coupled first-order differential equation (DE) system in the kinematic invariants. Inserting a Laurent expansion in the dimensional regulator $\epsilon$ and matching orders yields a hierarchy of coupled systems for the $\epsilon$-coefficients. With boundary conditions fixed using tools such as \textsc{AMFlow}~\cite{Liu:2022chg} or \texttt{pySecDec}~\cite{Borowka:2017idc}, these systems can, in principle, be solved numerically~\cite{Caffo:2002ch,Czakon:2008zk,Boughezal:2007ny,Mandal:2018cdj,Czakon:2021yub,Haisch:2024nzv,Czakon:2020vql,PetitRosas:2025xhm}. In practice this can be impractical: $\epsilon$-poles force computing high orders, singular phase-space make path choices non-trivial, and the complexity of the kinematic coefficients typically degrades numerical performance. Moreover, for sufficiently complicated multi-loop or multi-scale topologies, even obtaining the DEs can be challenging with current methods.\\

The analytical method first suggested in~\cite{Henn:2013pwa} is usually more efficient. By choosing a uniform-transcendental-weight basis that yields an $\epsilon$-factorised system, MI can be written as Chen iterated integrals. For canonical systems this implies computing polylogarithms~\cite{Goncharov:2010jf}, which can be organised via symbol technology and evaluated efficiently~\cite{Naterop:2019xaf}. However, practical and conceptual obstacles remain. For integral classes \emph{beyond} polylogarithms, where the DEs contain non-logarithmic kernels, closed forms require elliptic functions. Even in the case of logarithmic kernels, the appearance of multiple square roots can obstruct polylogarithmic representations and complicate robust numerical evaluation. In both cases, a pragmatic alternative is to avoid an analytic representation and instead solve via series-expansion methods~\cite{Lee:2017qql,Moriello:2019yhu}, implemented in public frameworks~\cite{Hidding:2020ytt,Armadillo:2022ugh,Prisco:2025wqs}, which give accurate results in regions covered by expansion points. In practice, one generates and interpolates numerical grids over the phase space. While very successful for low-dimensional kinematics, grid-based approaches suffer a severe curse of dimensionality, and they become less attractive for five-point (and higher-point) processes. This motivates numerical strategies that solve the DEs \emph{directly} along suitable paths in complexified kinematic space, aiming for stable and fast evaluations. In this context, the method of \textsc{PentagonFunctions}~\cite{Chicherin:2020oor,Chicherin:2021dyp,Abreu:2023rco} has been widely successful: it constructs a DE system for all special transcendental functions in the amplitude, evaluates them when closed forms exist, and otherwise computes the corresponding iterated integrals numerically. Although efficient, deriving analytic representations, and implementing iterated-integral machinery remains largely manual and non-automated. Moreover, it becomes suboptimal when the relevant kernels are not purely logarithmic. With recent progress in constructing an $\epsilon$-factorised basis from essentially any MI basis~\cite{e-collaboration:2025frv,Bree:2025tug}, non-logarithmic kernels are expected to appear more frequently, making the development of an optimal strategy for handling them crucial.\\

In this contribution, based on Ref.~\cite{PetitRosas:2025xhm}, we revisit numerical integration of the DEs for MI basis but merging the best aspects of the numerical and analytical approaches described above. We present a strategy tailored to genuinely multi-scale problems and show that it delivers substantial gains in speed and automation. The resulting framework targets execution times compatible with on-the-fly evaluation in Monte Carlo generators, while still enabling efficient parallel generation of numerical grids when such grids remain advantageous. The remainder of these proceedings is organised as follows. We first recall the DE setup for Feynman integrals and the role of canonical (polylogarithmic) forms, as well as extensions beyond polylogarithms. We then discuss numerical solution strategies, emphasising path selection and the analytic-continuation problem induced by algebraic branch cuts. Finally, we demonstrate performance on representative one- and two-loop five-point integral families relevant for phenomenological applications with multiple (possibly complex) kinematic scales, highlighting suitability for fast amplitude evaluation in event-generation workflows.

\section{The method of differential equations}
A scalar $l$-loop Feynman integral in dimensional regularisation can be written as
\begin{equation}
I^{d_0}_{a_1,\dots,a_N}
=\int \Big(\prod_{j=1}^{l}\mathrm{d}^{d_0-2\epsilon}k_j\Big)\,
\frac{1}{\prod_{i=1}^{N} D_i^{a_i}}\,,
\end{equation}
where $D_i$ are propagators encoding external kinematics and masses, $k_j$ are loop momenta, and the integer powers $a_i$ specify the integral. Integrals sharing the same set of propagators define a family and are related by IBPs identities; any topology reduces to a finite set of linearly independent integrals, the master integrals, whose choice is not unique. Given a basis vector $\vec I=(I_1,\ldots,I_n)$, differentiation with respect to a kinematic invariant $x$ followed by IBP reduction yields a closed system
\begin{equation}
\partial_x \vec I=\mathbb{A}(\vec x,\epsilon)\,\vec I\,,
\end{equation}
with $\mathbb{A}$ a matrix of rational functions in the invariants, $\vec{x}$, and in $\epsilon$. One can perform a set of analytical manipulations and redefine a new basis $\vec{J}$ such that the system becomes polynomial in $\epsilon$,
\begin{equation}
d\vec J=\sum_{k=0}^{\infty}\epsilon^k\,\mathbb{A}^{(k)}(\vec x)\,\vec J\,,
\end{equation}
where $\mathbb{B}$ may involve algebraic dependence on $\vec x$ (and $\epsilon$). With boundary data, this defines an initial value problem with formal solution
\begin{equation}
\vec J=\mathcal{P}\exp\!\left(\int_{\gamma}\sum_{k=0}^{\infty}\epsilon^k\,\mathbb{A}^{(k)}(\vec x)\right)\vec J(\gamma(0))\,,
\end{equation}
where $\gamma$ is a contour in complexified kinematics and $\mathcal{P}$ denotes path ordering.

\paragraph{Canonical and non-logarithmic forms:}
A further simplification occurs when the system can be cast in canonical form,
\begin{equation}\label{eq:canonical}
d\vec J=\epsilon\sum_{i=1}^{n}\mathcal{A}_i\,d\log\!\big(\alpha_i(\vec x)\big)\,\vec J\,,
\end{equation}
where the matrices $\mathcal{A}_i$ are constant (numerical) and the $\alpha_i$ are the \emph{letters} of the alphabet. Besides rational letters, alphabets often contain algebraic letters built from polynomials $P(\vec x),Q(\vec x)$ and square roots $r_i$, which can be written in Galois-manifest forms such as
\begin{align}
\alpha_i=\frac{P(\vec x)+Q(\vec x)\,r_k}{P(\vec x)-Q(\vec x)\,r_k}\,,\qquad
\alpha_i=\frac{P(\vec x)+Q(\vec x)\,r_k r_j}{P(\vec x)-Q(\vec x)\,r_k r_j}\,.
\end{align}
If a canonical form does not exist, we instead target $\epsilon$-factorised systems expressed through a basis of linearly independent one-forms,
\begin{equation}\label{eq:full}
d\vec J=\sum_{k=0}^{2}\epsilon^k\left[\sum_i \mathcal{A}_{k,i}\,d\log\!\big(\alpha_i(\vec x)\big)
+\sum_j \mathcal{B}_{k,j}\,\Omega_j(\vec x)\right]\vec{J}\,,
\end{equation}
where $\mathcal{A}_{k,i}$ and $\mathcal{B}_{k,j}$ are rational-number matrices and the $\Omega_j(\vec x)$ are non-logarithmic one-forms (typically rational functions possibly multiplied by a square root). Even though this has yet to be explored, the strategy we outline next also works for differential equations linear in $\epsilon$ that contain transcendental functions such as elliptic curves.

\paragraph{From master integrals to amplitude-level systems:}
Ultimately, we aim at numerical evaluations of scattering amplitudes rather than individual integrals. Following the amplitude-level strategy of~\cite{Caron-Huot:2014lda}, one can isolate differential equations for the special transcendental functions that actually enter the finite physical amplitude, yielding systems that are independent of $\epsilon$ and avoid computing redundant contributions. Concretely, by combining the $\epsilon$ expansion of $\vec J$ with the structure of the one-form decomposition above, we identify the physically relevant components and organise them by analytic structure using the symbol map~\cite{Goncharov:2010jf} along the lines of~\cite{Chicherin:2021dyp,Gehrmann:2024tds}. This organisation makes cancellations at the amplitude level systematic and removes spurious kernels that appear only in intermediate steps, thereby simplifying the final analytic structure and improving numerical stability. Even if we target to work with this type of system, the strategy is sufficiently general to also be applied to Eqs. \eqref{eq:canonical} and \eqref{eq:full}. 

\section{Numerical integration of differential equations}
The field of numerical integration of partial and total differential equations is vast, with a range of numerical techniques available to the individual. Selecting the correct method is key to an efficient integration. In our case, we aim at solving systems of the form of Eqs. \eqref{eq:canonical} and \eqref{eq:full}. which are both non-stiff and resolve to smooth solutions. We decide to focus on explicit methods to solve them and define our variables to be complex numbers, using the \texttt{Boost Odeint} \texttt{C++} library \cite{10.1063/1.3637934}. The integrating algorithm is chosen to be \texttt{bulirsch\_stoer} (BS), as it is specifically designed for very smooth solutions. To reduce CPU time, we exploit the structure of the differential equation: at each integration step we evaluate the analytic expressions for the one-forms and the relevant square roots once, and reuse them when assembling the right-hand side, avoiding repeated evaluation of common subexpressions. Since the $d\log$ letters and one-forms are rational functions (up to square-root factors) in the single evolving variable $x$, we precompute and store the polynomial coefficients in $x$ before evolving the variable, and reuse them throughout all steps along that segment. Their evaluation and compilation are further accelerated with the \texttt{FORM} optimiser~\cite{Kuipers:2013pba}; we observe substantial improvements in compilation time and CPU performance. An analogous caching strategy is applied to square roots: the roots of the underlying polynomials depend only on the non-evolving variables and are computed once per integration, and any square roots independent of the currently integrated variable are evaluated once and stored.\\

Reliable error control is essential because master integrals enter scattering amplitudes where numerical errors can be amplified or lead to cancellations. We account for (at least) two main error sources. 
\begin{enumerate}
    \item First, we control the integration error of the numerical algorithm itself by user-defined absolute and relative tolerances, $\mathcal{T}_A$ and $\mathcal{T}_R$, enforced by the controlled stepper in \texttt{Boost Odeint}. Thus, the error on the integrals evaluation is minimised. This is adequate for our use case, while leaving room for future refinements, such as minimising the global error on the amplitude instead, as used in~\cite{Badger:2025ljy}.
    \item Second, and often more critically, numerical instabilities can arise from singular structures in the differential equation itself. Letters and square roots introduce spurious and physical poles when they vanish. Both poles degrade numerical stability when the integration path approaches them, with sensitivity governed by the available precision and the singularity strength (denominator power). Spurious poles issues can in principle be handled via interpolation, whereas points near physical singularities may require higher precision or alternative treatments. In the present work we restrict ourselves to double and quadruple precision for the examples considered, implementing the latter via \texttt{GCC}'s \texttt{\_\_float128} type.
\end{enumerate}
  To further reduce errors, we use a dedicated routine to find paths designed to maintain sufficient distance from such singularities, except when the target point lies close to a pole.

\subsection{Finding a path}
\begin{figure}
    \centering
    \includegraphics[width=0.5\linewidth]{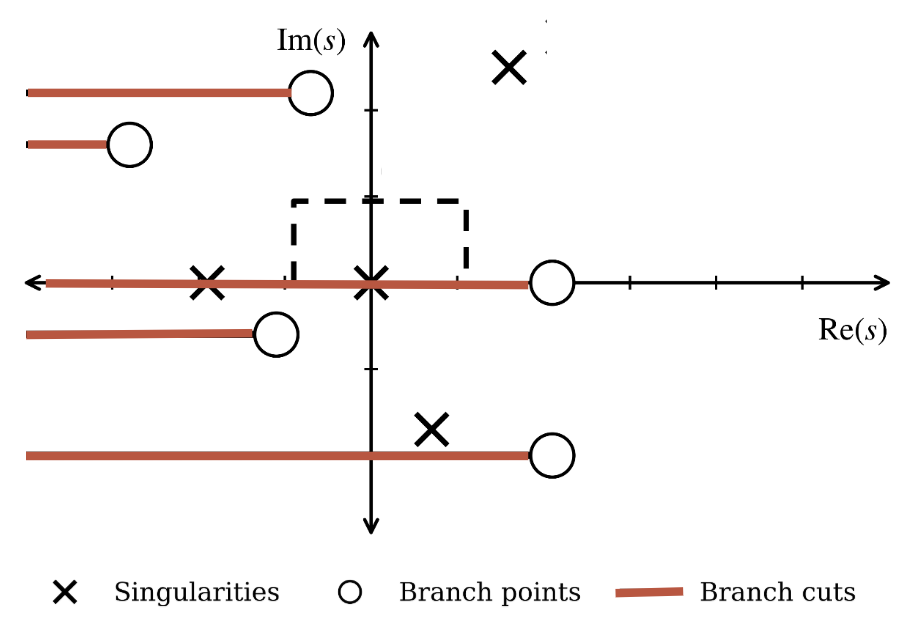}
    \caption{Representative complex space of $s$, with singularities, branch points and branch cuts of the system of differential equations, to be avoided in the numerical integration. The path, displayed as a dashed line, is composed of linear segments and designed to circumvent both singularities and branch cuts.}
    \label{fig:placeholder}
\end{figure}
We use {\tt AMFlow} to obtain a boundary condition at an arbitrary phase-space point, and evolve the differential equation from it one variable $x$ at a time, to the desired final point, following a path $\gamma(x,t_x) = (\lambda(x,t_x)_1,...\lambda_m(x,t_x))$ composed of $m$ linear segments $\lambda(x,t_x)$. A line parameter for each kinematic variable, $t_x$, is introduced to parametrise the segments, which only depend on $t_x$ itself and the kinematic variable $x \in \vec{x}$ that we are evolving at the time, such that
\begin{equation}
    \lambda(x,t_x) = x_0 + t_x (x_f-x_0)\,.
\end{equation}
To go from the origin $x_0$ to the desired final point $x_f$, we simply evolve $t_x$ from 0 to 1. The composition of the path in terms of linear segments allows one to circumvent the poles, identified analytically from the system of DEs,  by going into the complex plane.\\

There is an extra object that needs to be avoided. The square roots in the letters of the alphabet generate branch cuts that, if crossed, would make the numerical integration fail. To ensure we avoid them, we have to determine their location. Since we work with one kinematic variable at a time, we can decompose the multivariate polynomial inside the square roots in terms of its roots, precalculated at the start of each integration. The roots are also the branch cuts of the square root, which then reads 
\begin{align}
r_i(x) = \sqrt{\Pi^n_{j=1}\left( x-e_j \right)}\,,    
\end{align}
with $e_j$ being the root of the polynomial. We then work with the irreducible form, and choose the branch cuts to be parallel to the negative real axis. With this, we can always choose a path that avoids both singularities and branch cuts, as diagrammatically depicted in Figure~\ref{fig:placeholder}. By decomposing the square roots, we are free to rotate each branch cut individually, which is useful in certain cases. For example, when the cut has a small imaginary value, lying near the real axis, travelling below it would force us to approach singularities lying in the real axis. To avoid doing so, we rotate these cuts to be parallel to the imaginary axis. Other strategies could be used to deal with the branch cuts, such as not avoiding them and instead adjusting the sign of the square root, as used in~\cite{Badger:2025ljy}. 

\section{Results}
We show the power of the strategy by developing an integrator and applying it to two examples. All tests have been performed on a single thread of an \texttt{Intel(R)Core(TM) i7-13700H CPU @ 2.40 GHz}. First, we use the integrator to evaluate transcendental functions that appear in one-loop five-point amplitudes, coming from the Feynman integrals with 7 kinematic scales and up to transcendental weight 4, in particular, for the pentabox gauge invariant group of $e^+e^- \to \pi^+\pi^-\gamma$ up to $\mathcal{O}(\epsilon^2)$. Second, we implement two families, PB$_A$ and PB$_B$ of the two-loop process $p\bar{p} \to t\bar{t}+$jet, from~\cite{Badger:2024fgb}, and evolve up to transcendental weight four. \\

\begin{figure*}[t]
  \centering
  \makebox[\textwidth][c]{%
    \hspace*{-1cm}
    \includegraphics[width=0.55\textwidth]{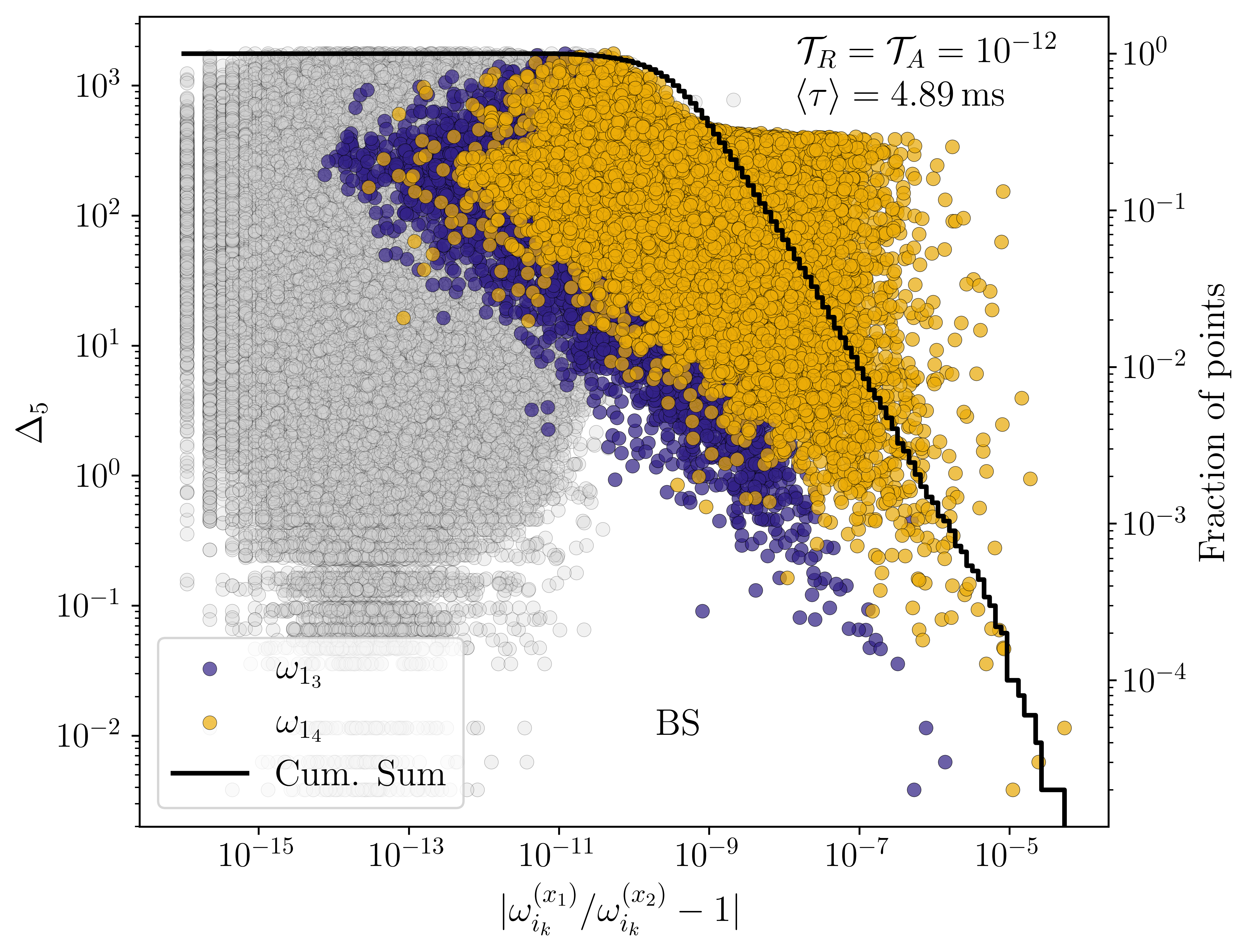}\hfill
    \includegraphics[width=0.55\textwidth]{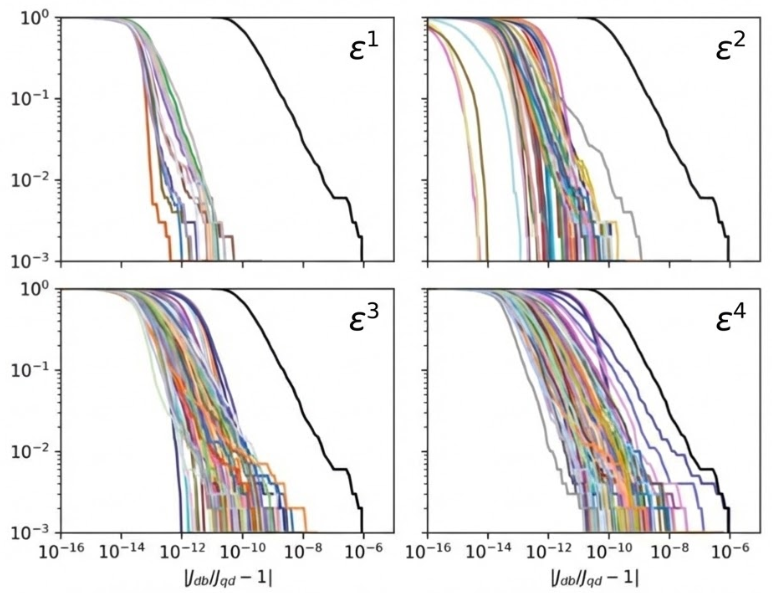}%
    \hspace*{-0.5cm}
  }
    \caption{\noindent\textbf{Left panel:} Relative discrepancy between the final values obtained from two different initial conditions as a function of the pentagon Gram determinant $\Delta_5$. The blue and yellow markers correspond, respectively, to $\omega_{1_3}$ and $\omega_{1_4}$, whereas the gray markers represent the remaining transcendental functions. The thick black curve reports, on the right vertical axis, the cumulative distribution of the relative differences. The average runtime, together with the absolute and relative tolerances employed, is indicated in the plot. \noindent\textbf{Right panel:} The cumulative distribution of the relative differences between the $\epsilon$-expansion coefficients of the PBb master integrals of Ref.~\cite{Badger:2024fgb}, computed at $10^3$ generic phase-space points using double versus quadruple precision. The black curve corresponds to the cumulative sum obtained after including all coefficients.}
    \label{fig:NN}
\end{figure*}

\paragraph{One-loop five-point massive process at $O(\epsilon^2)$:}We evaluate the DE system built in~\cite{PetitRosas:2025xhm}, with 65 transcendental functions, and estimate the error of the integrator by comparing the final values as obtained by starting from two different initial conditions. We generate a grid of 50k realistic phase-space points with the Monte Carlo event generator Phokhara~\cite{Rodrigo:2001kf,Campanario:2013uea,Campanario:2019mjh}, and plot the cumulative distribution in the left panel of Figure~\ref{fig:NN}, taking the relative precision value of the transcendental function with the worst error at each phase space point. We distinguish in the plot the more complicated transcendental functions of transcendental weight three and four, $\omega_{1_3}\text{ and } \omega_{1_4}$, that involve a pentagon integral, versus the pentagon Gram determinant $\Delta_5$. The plot showcases the good precision achieved by the integrator, while illustrating the fact that, as the final desired kinematics lie closer to a singularity, in this case $\Delta_5~=~0$, the numerical error increases. Furthermore, the average time $<\tau>$ is found to be of $\sim5$ ms per point.\\

\paragraph{Two-loop five-point integrals for $t\bar{t}j$:}We implement two families from~\cite{Badger:2024fgb} in our integrator, particularly PB$_A$ and PB$_B$. They involve six kinematic variables, and are composed of 88 and 121 master integrals, respectively, all needed at five orders in the dimensional regulator. For the first integral family, PB$_A$, a canonical form (see Eq. \eqref{eq:canonical}) is provided. This is different for PB$_B$, where instead an $\epsilon$-factorised form is provided with dependence on both logarithmic and one-forms, following Eq. \eqref{eq:full}. For each integral, we evolve each order of $\epsilon$ in the six kinematic variables from a boundary point to the provided benchmark value. Note that in a realistic implementation, there would be no need to evolve the centre-of-mass energy or the mass. We show on the right panel of Figure~\ref{fig:NN} the relative difference between double and quadruple precision for each MI $J$ of the PBb family. Each phase-space point takes around 0.1 seconds.

\section{Conclusions and outlook}
In this work, we presented a numerical framework for evaluating multi-scale and multi-loop Feynman integrals via differential equations, with particular emphasis on controlling their analytic structure. The method leverages three complementary formulations: canonical differential equations where the dimensional regulator $\epsilon$ factors out, representations in which the $\epsilon$ dependence is strictly polynomial, and systems for transcendental functions independent of the dimensional regulator. In all cases, algebraic functions are treated systematically through integration kernels cast as logarithmic forms or, more generally, differential one-forms. 

We validated the approach on demanding multi-scale examples, in particular on five-point one- and two-loop integrals with up to seven independent complex kinematic scales. For the amplitude application, organising transcendental functions by weight grading proved particularly beneficial, improving both efficiency and numerical stability. Finally, we addressed analytic continuation across kinematic regions with a new continuation strategy that evolves one variable at a time, yielding robust performance even for problems with many independent scales. \\

Future directions include extending our implementation to additional physical processes, with applications to $2\!\to\!2$ and $2\!\to\!3$~\cite{PetitRosas:2026kV} scattering channels already in progress. In parallel, we are carrying out a systematic benchmark against the \textsc{PentagonFunctions} approach, where preliminary tests indicate encouraging improvements in runtime. A further natural, and likely necessary, development is to generalise the strategy to elliptic differential equations. Together with recent advances in constructing $\epsilon$-factorised systems, this extension is expected to broaden the range of processes that can be computed efficiently and reliably within our framework.
\acknowledgments
I wish to thank V. Sotnikov for interest and feedback on the project. This work was supported by the Leverhulme Trust, LIP-2021-014.

\bibliographystyle{JHEP}
\bibliography{biblio}

\end{document}